\documentclass[%
 reprint,
 superscriptaddress,
 floatfix, 
 amsmath,amssymb,
 aps,
 twocols
]{revtex4-1}

\usepackage{graphicx}
\usepackage{dcolumn}
\usepackage{bm}
\usepackage{color}
\usepackage[caption=false]{subfig}
\usepackage{listings} 
\usepackage{amsmath}   
\usepackage{array}
\usepackage{colortbl}
\usepackage{xcolor,colortbl}
\usepackage{textgreek}
\usepackage{float}
\usepackage{mathrsfs}  
\usepackage{amsmath,amssymb}  
\usepackage{siunitx} 

\usepackage{array,ragged2e}

\usepackage{booktabs,amsmath,caption}

\usepackage{makecell}

\usepackage{adjustbox}
\usepackage{multirow}

\begin{document}

\newcommand{\cfbox}[2]{%
    \colorlet{currentcolor}{.}%
    {\color{#1}%
    \fbox{\color{currentcolor}#2}}%
}

\title{Ligand-protein interactions in lysozyme investigated through a dual-resolution model}


\author{Raffaele Fiorentini}
\affiliation{Max Planck Institute for Polymer Research, Mainz, Germany}

\author{Kurt Kremer}
\affiliation{Max Planck Institute for Polymer Research, Mainz, Germany}

\author{Raffaello Potestio}
\email{raffaello.potestio@unitn.it}
\affiliation{Physics Department, University of Trento, via Sommarive, 14 I-38123 Trento, Italy}
\affiliation{INFN-TIFPA, Trento Institute for Fundamental Physics and Applications, I-38123 Trento, Italy}


\date{\today}

\begin{abstract}
A fully atomistic modelling of biological macromolecules at relevant length- and time-scales is often cumbersome or not even desirable, both in terms of computational effort required and {\it a posteriori} analysis. This difficulty can be overcome with the use of multi-resolution models, in which different regions of the same system are concurrently described at different levels of detail. In enzymes, computationally expensive atomistic detail is crucial in the modelling of the active site in order to capture e.g. the chemically subtle process of ligand binding. In contrast, important yet more collective properties of the remainder of the protein can be reproduced with a coarser description. In the present work, we demonstrate the effectiveness of this approach through the calculation of the binding free energy of hen egg white lysozyme (HEWL) with the inhibitor di-N-acetylchitotriose. Particular attention is posed to the impact of the mapping, i.e. the selection of atomistic and coarse-grained residues, on the binding free energy. It is shown that, in spite of small variations of the binding free energy with respect to the active site resolution, the separate contributions coming from different energetic terms (such as electrostatic and van der Waals interactions) manifest a stronger dependence on the mapping, thus pointing to the existence of an optimal level of intermediate resolution.
\end{abstract}

\maketitle

\section{Introduction}\label{sect:intro}

One of the most relevant challenges of computational biochemistry and biophysics is the accurate calculation of binding free energies \cite{FE1, FE2, FE3}, which represents one of the key steps in the identification of pharmacological targets as well as in the development of new drugs \cite{drug1, drug2, drug3}. However, the large sizes of the molecules under examination (often above the hundred of residues), as well as the necessity to screen through large datasets of potential candidate molecules, make this effort onerous in terms of time and computational resources.

A promising way to mitigate these limitations is the use of multiple-resolution models of the protein, that is, representations in which different parts of the molecule are concurrently described at different levels of resolution \cite{mult2,adress1, adress2,mult1, mult3,mult4,mult5,mult6,Kremer_Proteins_2016-lys_multires}. The chemically relevant part of the protein, e.g. the active site, is modelled at level of detail, typically atomistic. For the remainder, on the contrary, a simplified representation is used, where several atoms are lumped together in effective interaction sites. The working hypothesis underlying these methods is that only a relatively small part of the molecule requires an explicitly atomistic treatment; the remainder, in fact, is mainly responsible for large-scale, collective fluctuations whose function-oriented role is well recognised and prominent \cite{Amadei1993,Carnevale2006,Zen2007,Pontiggia2008,Kremer_Proteins_2016-lys_multires}, however also prone to be accurately reproduced by lower-resolution representations \cite{Tirion_ENM,Hinsen1998,Delarue2002,Micheletti2004,Romo_et_al,Potestio2009}. Hence, the resulting model favourably joins the accuracy of an atomistic (AT) description where needed and the computational efficiency of a coarse-grained (CG) one where possible.

In order to take full advantage of the dual-resolution approach to protein modelling, though, one has to solve a few key open issues: first, the definition of the appropriate coarse-grained model to employ in the low-resolution part \cite{Gohlke2006,ZHANG_BJ_2008,ZHANG_BJ_2009,Potestio2009,ZHANG_JCTC_2010,Sinitskiy2012,Polles2013,Foley2015,digginsJCTC2018}; second, the coupling between high- and low-resolution models, which has to be performed so as to guarantee that the appropriate observables are reproduced with respect to the reference provided for example by a fully atomistic simulation. This issue entails a further one, namely the identification of the correct observables apt to quantify the fidelity with which the behaviour of the system is reproduced by the dual-resolution model; third, the selection of the subpart of the molecule that {\it requires} a high-resolution modelling. In the present work we will focus specifically on this third aspect.

Various methods and approaches have been developed in the past few years to describe proteins in dual resolution \cite{mult1, mult2, mult3, mult4, mult5, mult6}. In general, the high-resolution part is modelled at the all-atom level, making use of one of the several atomistic force fields available. The coarse-grained representations range from simple bead-spring elastic networks \cite{Tirion_ENM,Micheletti2004,Kremer_Proteins_2016-lys_multires} to more sophisticated G\=o-type models \cite{mult3}. Recently, we have proposed a dual-resolution model \cite{Kremer_Proteins_2016-lys_multires} where, in the CG part, only the C$_\alpha$ carbons of the protein chain are retained and connected one with the other by harmonic bonds. This model has been employed in the present work with the aim of assessing the accuracy of a hybrid atomistic/coarse-grained description of a protein for binding free energy calculations. The system under examination is hen egg-white lysozyme in explicit water, bound to a sugar substrate, di-N-acetylchitotriose. We carried out calculations of the binding free energy of the ligand in the active site, with a twofold objective. In fact, not only we aimed at verifying that the computed quantity in the dual-resolution model matches a reference, all-atom calculation; but rather we also investigated the impact of different choices in the definition of the high-resolution subdomain. This aspect bears the highest prominence, as it is becoming increasingly more evident that a crucial component in the construction of accurate and effective low-resolution models for biological and soft matter systems is represented by the mapping \cite{Foley2015,Kremer_Proteins_2016-lys_multires,digginsJCTC2018}, that is, the particular selection of collective variables employed to describe the system. Here, we provide novel evidence of this general property in the context of a dual-resolution model of a biomolecule, and describe a transferable strategy to tackle this issue.

\section{Methods}\label{sect:material-methods}

The system under examination in the present work is hen egg-white lysozyme (HEWL) in aqueous solution. In this model, the binding site of the enzyme and the substrate molecule, the inhibitor di-N-acetylchitotriose, are represented with atomistic detail. The protein model employed is not adaptive, that is, the resolution of a given residue is fixed --either atomistic or coarse-grained-- and does not change throughout a simulation. However, at difference with other works \cite{mult3, adress1, adress2}, several values of the number of protein residues treated at high resolution have been explored and employed in independent calculations. The impact of choosing different numbers of active site residues to model at the atomistic level is a central aspect of this study. The coarse-grained model employed to describe the low-resolution part of the protein is a simple bead-spring representation where the selected sites (namely the C$_\alpha$ atoms) are connected by elastic bonds penalising the deviations from the distances that interacting atoms have in the reference conformation. Two values of elastic constants employed, one for C$_\alpha$'s along the chain, and one for all other bonds. Water molecules are described in atomistic detail throughout the whole simulation box: the interaction with the high-resolution part of the protein takes place through the standard all-atom force field, while the interaction with the coarse-grained beads is mediated by a purely repulsive potential acting on the sole oxygen atom. 

Hereafter we provide a detailed description of the model. We first discuss the calculation of the binding free energy $\Delta G_{bind}$, then we outline the dual-resolution model and its coupling to the atomistic part, and finally report information about the simulation setup. Further details are made available in the \textit{Supporting Information}.

\subsection{Binding Free Energy calculation} \label{subsect:binding-FE}

One of the key points of this work is the calculation of the protein-ligand binding free energy $\Delta G_{bind}$, which quantifies the affinity of a molecule towards a protein \cite{FE1, FE2, FE3}. As such, it plays a prominent role in the investigation of the biochemical function and activity of enzymes and similar biomolecules, and in the development of effective drugs.

$\Delta G_{bind}$ is defined as the difference between the free energy of the system in the configuration in which the ligand is bound to the active site ($G_{b}$) and the corresponding value when the ligand is absent ($G_{ub}$):

\begin{equation}
\Delta G_{bind} =  G_{b} - G_{ub}
\end{equation}
\medskip

This value, in the specific case under examination, varies according to the number of active site residues modelled with atomistic resolution, as we will see in Sect. \ref{res-disc}.

The free energy difference between two states is here computed by means of thermodynamic integration (TI) \cite{TI}. Specifically, a scalar $\lambda \in [0,1]$ is defined which parametrises the potential energy of the system as $U_\lambda({\bf r}) = \lambda U_{A}({\bf r}) + (1 - \lambda) U_{B}({\bf r})$ connecting the states A and B. The sought quantity is given by:

\begin{equation}
\Delta G =  \int _{0}^{1}\left\langle {\frac {\partial U(\lambda )}{\partial \lambda }}\right\rangle _{\lambda }d\lambda
\end{equation}
\medskip

\begin{figure}[h]
\begin{center}
\includegraphics[width=\columnwidth]{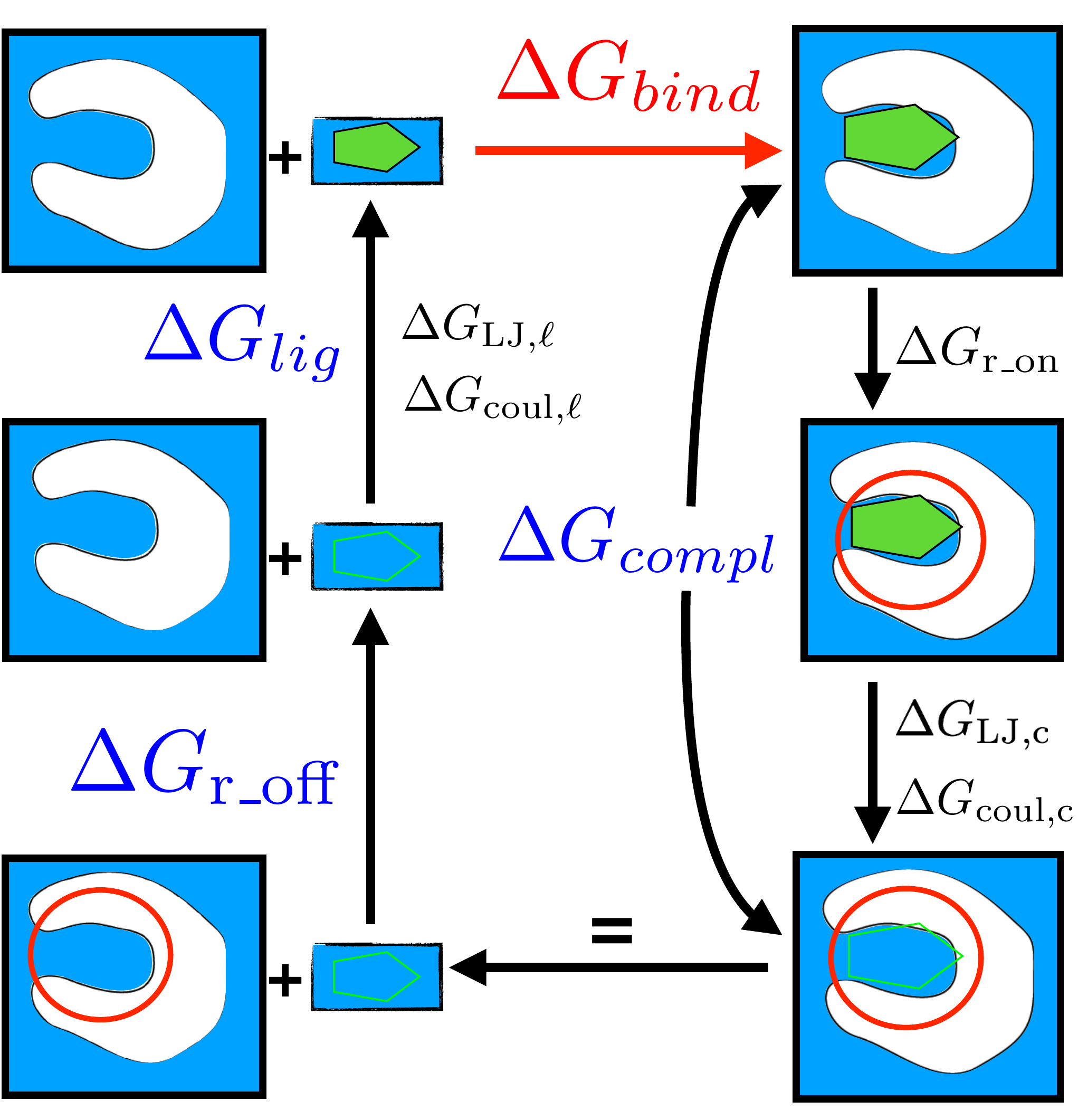}
\end{center}
\caption{pictorial representation of thermodynamic cycle. Starting from the top-right corner of the figure, we decouple the ligand from the protein $ (\Delta G_{compl}$, which also includes a set of restraints between ligand and protein) and subsequently introduce it in water ($\Delta G_{lig}$). A further step is the restraints removal ($\Delta G_{r\_off}$) whose calculation is analytical.}
\label{fig:th-cycle-myself}
\end{figure}

Since the free energy is a state function, the nature of the path is unimportant, and one can choose a thermodynamic cycle that connects the bound and unbound states through several intermediate ones, as illustrated in Fig.\ref{fig:th-cycle-myself}. In particular, we can identify two main terms: the insertion of the ligand from vacuum to water $\Delta G_{lig}$, and the decoupling from the protein $\Delta G_{compl}$. A further step is the removal of the restraints that keep the ligand in proximity of the protein during the damping of the ligand-protein interactions, $\Delta G_{r\_off}$; this latter calculation can be carried out analytically without the need to run simulations. Hence, $\Delta G_{bind}$ is the algebraic sum of the previous three terms:
\begin{equation}
\Delta G_{bind} = \Delta G_{compl} + \Delta G_{lig} + \Delta G_{r\_off}
\end{equation}

According to the previous definitions of each term, neither $\Delta G_{lig}$ nor $\Delta G_{r\_off}$ changes with the protein resolution: indeed, the former corresponds to the solvation free energy of the ligand, which is always treated at the atomistic level; likewise, the calculation of the restraint removal free energy is analytic \cite{FE3}. The unique term that varies depending on the number of active site residues modelled in high resolution is the free energy change of the protein-ligand complex between the bound state and the state where the ligand is removed, that is, the variation of $\Delta G_{bind}$ is equal to the variation of $\Delta G_{compl}$.

The alchemical change in the calculation of $\Delta G_{compl}$ is performed in three steps (in the following, the subscripts $c$ and $\ell$ stand for complex and ligand, respectively). First, one adds a set of restraints between protein and ligand ($\Delta G_{r\_on}$) in order to avoid the problem of the ligand leaving the binding pocket when interactions are being removed. The presence of restraints is indicated in the cycle scheme of Fig.\ref{fig:th-cycle-myself} with a red circle: it represents the fact that the ligand is confined in a certain volume. For this work we use the set of restraints described by Boresch \cite{FE3}. Second, Coulomb interactions are switched off ($\Delta G_{coul,c}$); third, the Lennard-Jones potentials modelling van der Waals interactions are removed ($\Delta G_{LJ,c}$). Likewise, the alchemical change in the ligand free energy $\Delta G_{lig}$ is performed in two steps: first switching on Coulomb interaction ($\Delta G_{coul,\ell}$), and then Lennard-Jones ($\Delta G_{LJ,\ell}$).
The last contribution to the binding free energy, $\Delta G_{r\_off}$, derives from restraint removal: its calculation is analytical and therefore it does not require alchemical changes. These transformations are summarised in Fig. \ref{fig:th-cycle-myself} and Tab. \ref{tab:alch-chang}. Further details can be found in the \textit{Supporting Information} in the section relative to the thermodynamic cycle.

\begin{table}[pbth]
\setlength{\tabcolsep}{0mm}
\caption{Summary of the alchemical changes and the protein resolution dependence for each contribute of Binding free energy $\Delta G_{bind}$.}
\label{tab:alch-chang}
\centering
\begin{tabular}{|l 
               |c|c|}
\hline
\multicolumn{1}{|l|}{}  &  &   \multicolumn{1}{c|}{\bfseries prot. res.} \\
 & \multirow{-2}{*}{\bfseries alchemical changes} & \multicolumn{1}{c|}{\bfseries dependence}  \\   %
\hline
\textbf{\textDelta G\textsubscript{compl   }}  & $\Delta G_{\text{coul,c}}$ + $\Delta G_{\text{LJ,c}}$ + $\Delta G_{\text{r\_on}}$                   & YES \\ 
\hline
\textbf{\textDelta G\textsubscript{lig   }}      & \multicolumn{1}{l|}{$\Delta G_{\text{coul},\ell}$ + $\Delta G_{\text{LJ},\ell}$}                   & NO \\ 
\hline
\textbf{\textDelta G\textsubscript{r\_off   }}      & \multicolumn{1}{l|} {Analytical}  & NO \\ \hline    
\end{tabular}
\end{table}

The calculation of $\Delta G_{compl}$ can be carried out in two different ways, namely decoupling and annihilation. Decoupling refers to turning off the interaction between the molecule and its environment, while maintaining the potentials among atoms constituting the molecule; annihilation, on the other hand, implies turning off the interaction between the molecule and the environment {\it as well as} the intramolecular interaction. Here we consider the values of $\Delta G$ obtained through ligand decoupling, since this process is more intuitive with respect to annihilation; furthermore, the ligand is always treated at fully atomistic detail, therefore it is not involved in the change of free energy while varying the protein resolution. In Tab. \ref{tab:finalres} and Fig. \ref{fig: comparison} (and with greater detail in the \textit{Supporting Information}, annihilation section) we provide data showing that the values of binding free energy obtained using decoupling and annihilation are consistent within the error bars.

\subsection{Dual-Resolution protein model}\label{Dual-res-model}

\begin{figure}[htp]
\begin{center}
\includegraphics[width=\columnwidth]{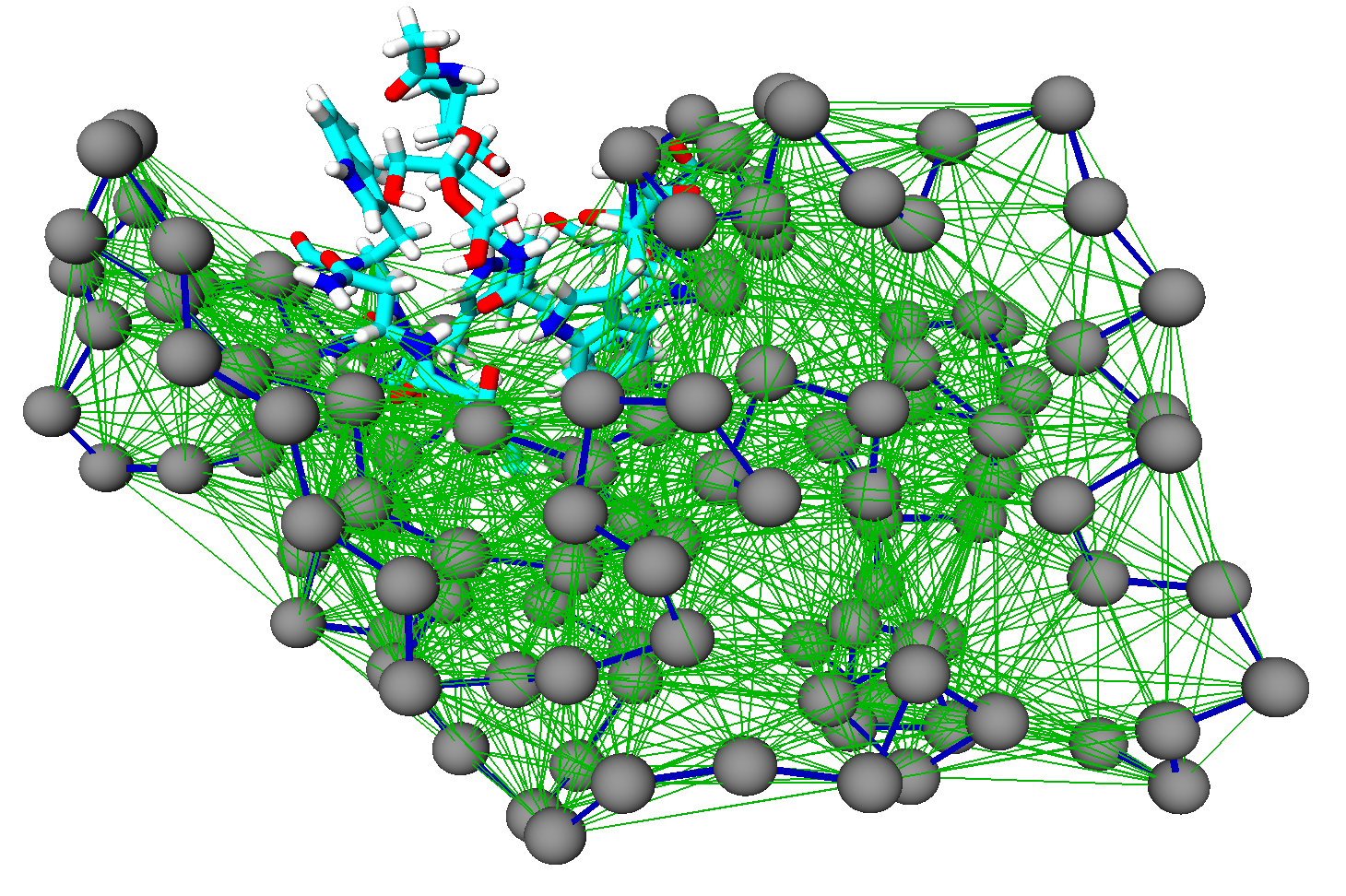}
\end{center}
\caption{Visualisation of the dual-resolution protein \cite{Kremer_Proteins_2016-lys_multires}. The residues included in atomistic detail are shown in red, blue, cyan and white (O, N, C and H atoms). The grey spheres are ENM nodes, the stiff backbone springs are shown as dark blue lines and all others (weaker) springs are shown in green.}
\label{fig:prot-lig}
\end{figure}

In this work the solvent is treated with all-atom detail, while the protein has a fixed (i.e. position- and time-independent) dual-resolution. The binding site is modelled with atomistic resolution, whereas the rest of the protein is coarse-grained. To describe the lower-resolution part we employ an elastic network model (ENM) \cite{Tirion_ENM, Kremer_Proteins_2016-lys_multires}, in which each residue is mapped onto a bead whose position corresponds to the $C_\alpha$ atom in the atomistic description. These beads are connected by harmonic springs as shown in Fig.~\ref{fig:prot-lig}.

The potential energy is given by: 
\begin{equation}
E = \sum_{i}\sum_{j}k_{ij}\left(r_{ij} - r_{ij}^0\right)^2 \theta(r_c-r_{ij}^0)
\end{equation}
\medskip
with spring constants $k_{ij}$, equilibrium distance $r_{ij}^0$, a cutoff distance $r_{c}$, $i$ and $j$ are the node index, and $\theta(r)$ is a Heaviside theta function taking value $1$ if $r>0$ and $0$ otherwise. In this model we made use of two different elastic constants: a very stiff spring ($k_b$) for consecutive beads, represented in blue in Fig.~\ref{fig:prot-lig}; and a weaker spring $k_{nb}$ for not consecutive beads whose distance in the reference (native) conformation lies below a fixed cutoff (in green).

The ENM used here is parametrised to reproduce the conformational fluctuations of the reference all-atom model, these being quantified by the root mean square fluctuations (RMSF) of the all $C_{\alpha}$ atoms of the system \cite{Kremer_Proteins_2016-lys_multires}. The residues in direct contact (H-bonding or hydrophobic contact) with the substrate are modelled with all-atom detail; in order to select the other binding site residues to be described at the atomistic level, we sorted them by increasing distance of their the center of mass from the closest ligand atom.

The water-CG protein interaction consists in a simple excluded volume, modelled {\it via} a Weeks-Chandler-Anderson (WCA) potential \cite{WCA}. The details about the procedure followed to determine the ENM elastic constants and the excluded volume interaction are provided in the \textit{Supporting Information}, while the numerical values of the resulting parameters are reported hereafter. 

\subsection{Simulation details}\label{sim-details}

The reference model is given by the 2 ns equilibrated PDB structure 1HEW in the NPT ensemble (the Parrinello-Rahman barostat \cite{Parrinello} with a time constant of 2.0 ps and 1 bar was used). Both fully atomistic and dual-resolution models of HEWL are solvated in water and placed in a cubic simulation box of 7.06 nm side. The force field employed is Amber99SB \cite{amber99sb-1}, whereas the water model is TIP3P \cite{tip3p}.  The inhibitor, which was always atomistic, had GLYCAM forcefield parameters consistent with Amber99SB \cite{amber99sb-2}. The TI binding free energy calculation consists of 3 different steps: $\Delta G_{compl}$, $\Delta G_{r\_off}$, $\Delta G_{lig}$:

\begin{enumerate}
\item{The protein-ligand complex free energy (\textDelta G\textsubscript{compl}) calculation uses 11 $\lambda$ values per $\Delta G_{restr\_on,c}$, 5 evenly spaced $\lambda$ values per $\Delta G_{LJ,c}$ (with separation 0.20) and 15 $\lambda$ values per $\Delta G_{coul,c}$, with 600 ps of simulation per $\lambda$ in the fully atomistic case, and 4000 ps in the dual-resolution case to improve the statistics.}
\item{The restraint removal free energy (\textDelta G\textsubscript{r\_off}) calculation is analytical (details on \textit{Supporting Information}).}
\item{The ligand solvation free energy (\textDelta G\textsubscript{lig}) calculation uses 5 evenly spaced $\lambda$ values per $\Delta G_{coul,\ell}$ (with separation 0.20) and 16 $\lambda$ values per $\Delta G_{LJ,\ell}$, with 600 ps of simulation of each $\lambda$-value.  }
\end{enumerate}

In the thermodynamic integration we employ the soft-core potential of Ref. \cite{manual_grom} with parameters $\alpha = 0.5$ and $p = 1.0$ to avoid possible singularities in the Lennard-Jones terms from atoms overlapping during the alchemical change.  The temperature is kept constant at 298 K by means of a Langevin thermostat with a friction constant $\gamma = 15$ $ps^{-1}$. The integration step is 1 fs. The calculation of electrostatic interaction is performed using the reaction field method with a dielectric constant $\epsilon = 80$ and a cutoff of 1.2 nm. These parameters are a good compromise between speed and accuracy, as verified in Ref. \cite{Shirts}. The SETTLE \cite{settle} and RATTLE \cite{rattle} algorithms for rigid water and rigid bonds to hydrogen have been used. Each system is prepared using fully atomistic minimisation with steepest descent and 6 ns of equilibration in NVT (for both ligand-free and ligand-bound systems). All simulations (both fully atomistic and dual-resolution) are carried out with the ESPResSo++ simulation package \cite{espp1, espp2}, in which we have implemented TI (except in case of annihilation, for which all steps are performed in both ESPResSo++ and GROMACS \cite{grom}). Some preliminary fully atomistic equilibration simulations use GROMACS. The error bars shown are calculated using the Student $t$ at 95\% confidence limit \cite{student}, via standard deviations obtained using block averaging in which all trajectories are divided into four blocks of equal length. 

The parametrization of the dual-resolution model is consistent with the work in Ref.\cite{Kremer_Proteins_2016-lys_multires}: the spring constant between consecutive $C_{\alpha}$ nodes along the backbone ($k_b$) has a stiff value of $5 \cdot 10^4$ $kJ \cdot mol^{-1} \cdot nm^{-2}$, whilst all the other ones ($k_{nb}$) have a value of  $160$ $kJ \cdot mol^{-1} \cdot nm^{-2}$, until 1.2 nm as cutoff, parametrised by minimising the average root mean square error in the $C_{\alpha}$ RMSF.  
Moreover, a WCA interaction is applied between $C_{\alpha}$ nodes and all solvent molecules center of mass. In the WCA potential, $\epsilon$ has a value of $0.34 \hspace{2mm} kJ \cdot mol^{-1}$ arbitrarily chosen as the value for carbon in the atomistic forcefield, and $\sigma_{i} = R_{g,i} \cdot c$ where $R_{g,i}$ is the radius of gyration of a given residue $i$ where $c$ is the same for all amino acids. The value of $c$ is tuned to give the correct bulk water density of reference for a protein-water system. The $c$ value found is 0.658. Further explanations about $c$ can be found in the \textit{Supporting Information}.

\section{Results and discussion}\label{res-disc}

We performed the calculation of $\Delta G_{b}$ of lysozyme modelled in dual-resolution, varying the number of atomistic residues constituting the binding site and comparing the results with a fully atomistic reference simulation. Recall that the binding free energy calculation consists of three steps: restraint removal, ligand $\Delta G$, and ligand-complex $\Delta G$; of these, only the latter depends on protein resolution, that is, only $\Delta G_{compl}$ assumes different values for different numbers of active site residues described at the all-atom level.

As explained in the previous section, the contribution coming from the restraints can be analytically computed and amounts to $\Delta G_{r\_off} = -31.3 \hspace{2mm} kJ \cdot mol^{-1}$. Likewise, the Coulomb and Lennard-Jones contributions to the ligand free energy $\Delta G_{lig}$ are the following: 

\begin{equation*}
 \begin{split}
&\Delta G_{coul,\ell} = -142.8 \pm 1.7 \hspace{2mm} kJ \cdot mol^{-1} \\
& \Delta G_{LJ,\ell} \hspace{1.8mm}= -9.1 \pm 6.3 \hspace{2mm} kJ \cdot mol^{-1} \\
 \end{split}
\end{equation*}

Hence: 

\begin{equation*}
\Delta G_{lig} = -151.9 \pm 8.0 \hspace{2mm} kJ \cdot mol^{-1}  
\end{equation*}

\begin{figure*}[t!]
\subfloat[]{\includegraphics[width=\columnwidth]{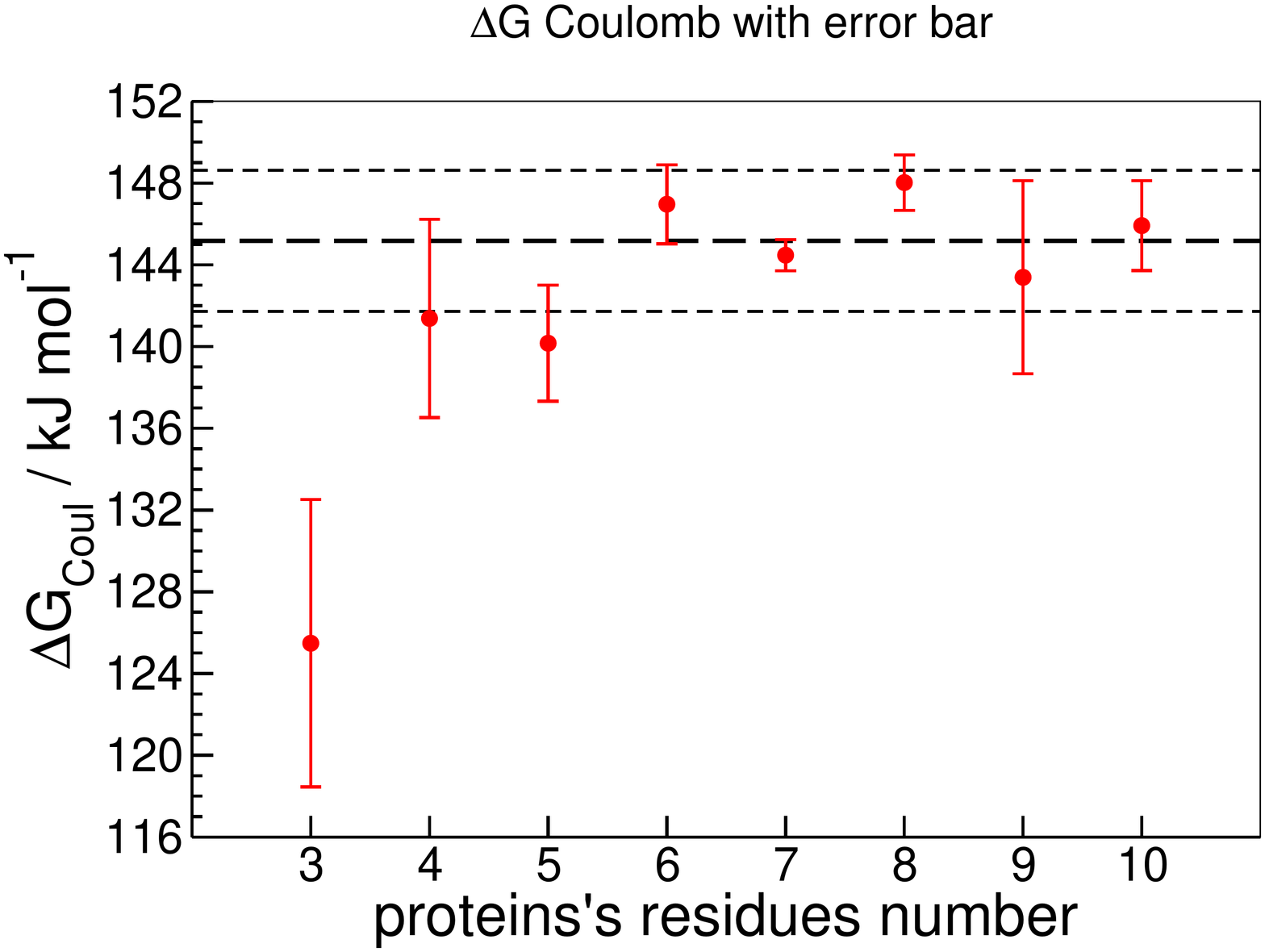}}
\subfloat[]{\includegraphics[width=\columnwidth]{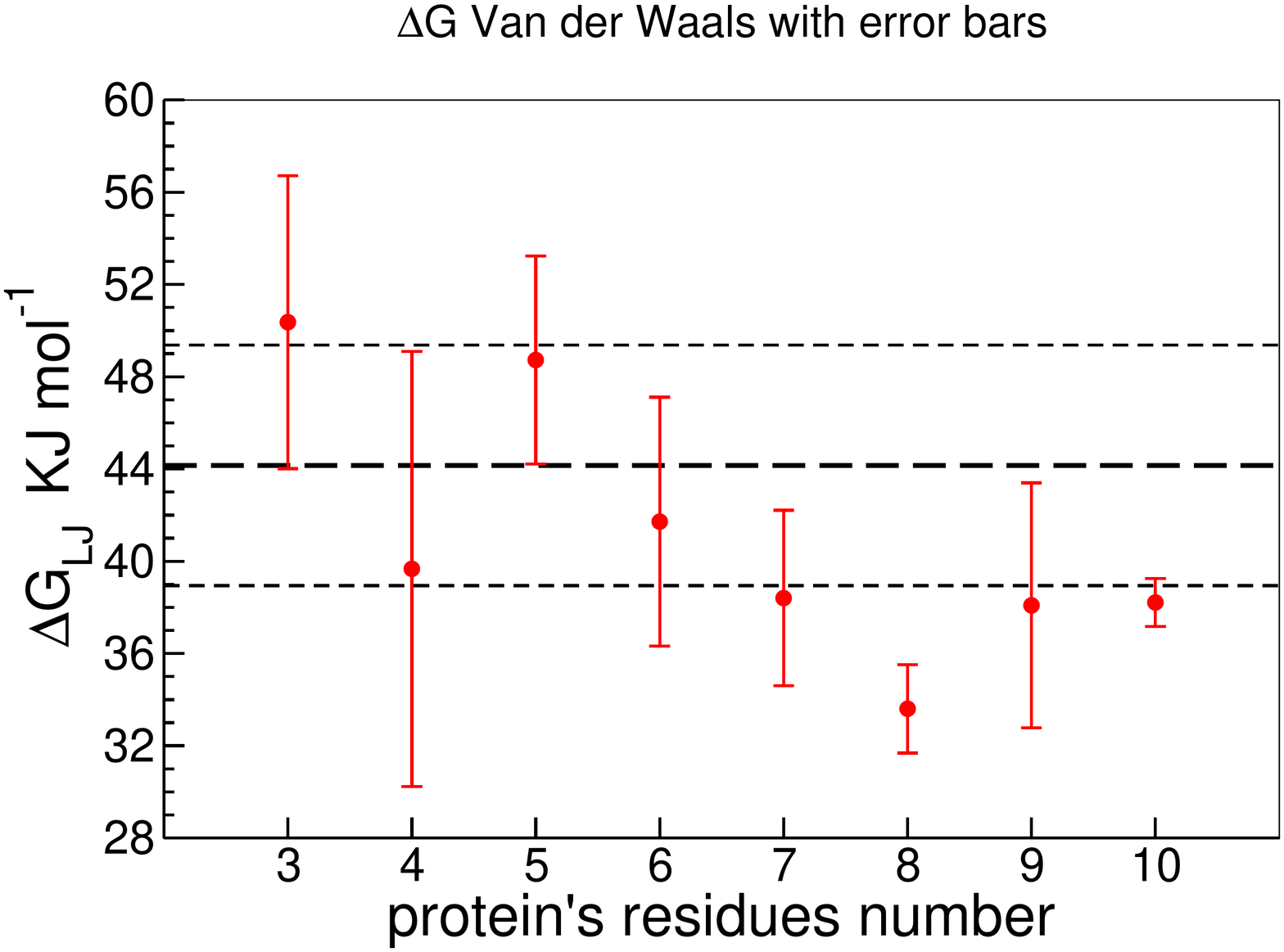}}\\
\subfloat[]{\includegraphics[width=\columnwidth]{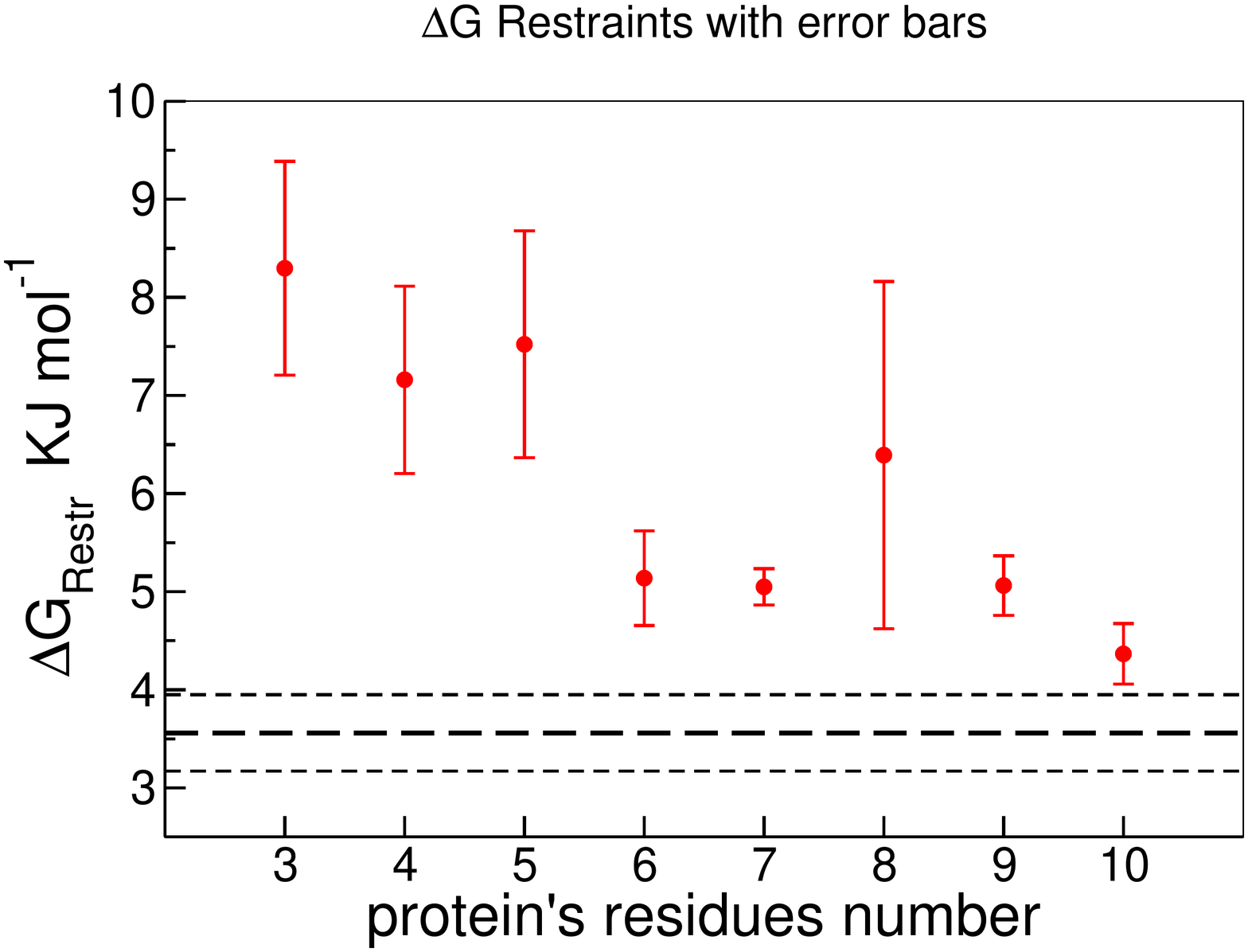}}
\subfloat[]{\includegraphics[width=\columnwidth]{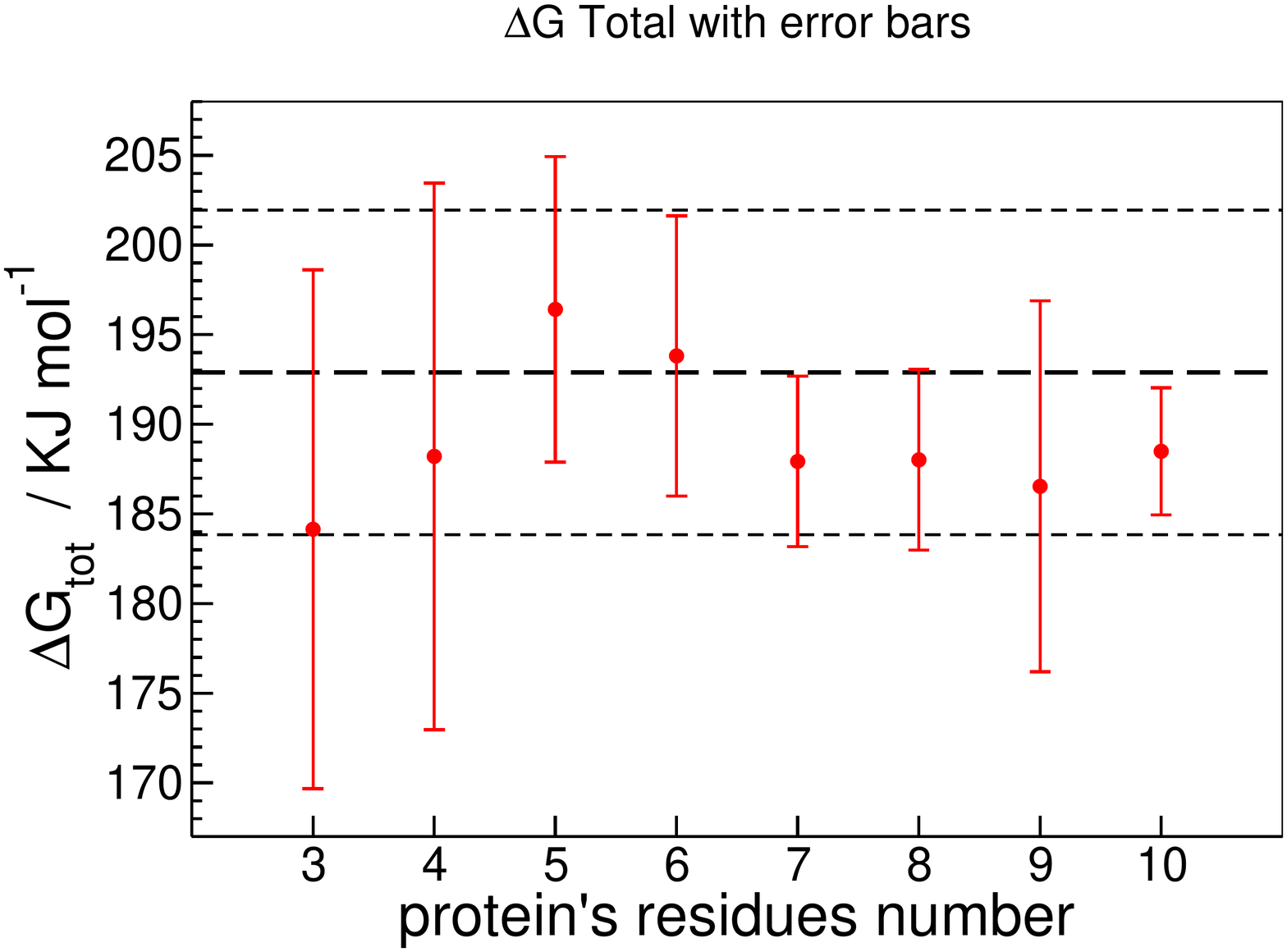}}
\caption{(a) Coulomb, (b) Lennard-Jones, (c) restraint and (d) total free energies in the protein-ligand complex, as a function of protein's residues number included in atomistic detail in the multi-resolution set-up. The heavy dashed black horizontal lines are the reference values from fully atomistic simulations, and the lighter dotted black horizontal lines are the error bars for those values. These simulations use decoupling, not annihilation. y-axes do not cover the same energy range.}
\label{fig:_at_vs_dualres_fn_aa}
\end{figure*}

The final step is the calculation of $\Delta G_{compl}$, whose results, including the comparison between dual-resolution model and fully atomistic reference, are shown in Tab.~\ref{tab:free-energy} and illustrated in Fig.~\ref{fig:_at_vs_dualres_fn_aa}.

\begin{table*}[t!]
\setlength{\tabcolsep}{0mm}
\centering
\caption{In this table are reported the resulting values of free energy of Complex Free Energy (4th column) and its components (Coulomb, Lennard Jones and Restraints respectively in the first three columns) in fully atomistic system and varying the number of atomistic residues. All the values are in $kJ \cdot mol^{-1}$ and performed with Thermodynamic Integration. Moreover, all simulations are carried out in ESPResSo++. In particular, for each value of $\lambda$, the dual-resolution simulations with different number of atomistic residues last 4 nsec; the atomistic simulation, instead, lasts 0.6 ns (600 ps)}
\label{tab:free-energy}
\begin{tabular}{@{}|l|
                @{\hspace*{0mm}} S[table-format=4.1(3)] |
                @{\hspace*{0mm}} S[table-format=3.1(3)] |
                @{\hspace*{0mm}}  S[table-format=2.1(3)] |
                @{\hspace*{0mm}}  S[table-format=4.1(4)] | 
                @{}}
\hline
\textbf{at res}  & \textbf{\textDelta G\textsubscript{Coul,c}} & \textbf{\textDelta G\textsubscript{LJ,c}} & \textbf{\textDelta G\textsubscript{Restr\_on,c}} & \textbf{\textDelta G\textsubscript{compl}} \\
\hline
\text{fully-at   } & 145.2 \pm 3.5  & 44.2 \pm 5.2 & 3.6 \pm 0.4 & 193.0 \pm 9.1 \\ \cline{1-5}
\hline
aa-3 &  125.5 \pm  7.0    & 50.4 \pm 6.3  & 8.3 \pm 1.1   & 184.2 \pm 14.4    \\ \cline{1-5}
\hline
aa-4 &  141.4 \pm 4.9    & 39.7 \pm 9.4  & 7.2 \pm 1.0   & 188.3 \pm 15.3   \\ \cline{1-5}
aa-5 &  140.2 \pm 2.8    & 48.7 \pm 4.5 & 7.5 \pm 1.2   & 196.4 \pm 8.5 \\ \cline{1-5}
\hline
aa-6 & 147.0  \pm 1.9     & 41.7 \pm   5.4  & 5.1 \pm  0.5  & 193.8 \pm   7.8 \\ \cline{1-5} 
\hline
aa-7 & 144.5  \pm 0.8   & 38.4 \pm  3.8  & 5.0 \pm  0.2  & 187.9 \pm   4.8 \\ \cline{1-5} 
\hline
aa-8 & 148.0 \pm 1.4   &  33.6 \pm 1.9 & 6.4 \pm 1.8  &  188.0 \pm  5.1 \\ \cline{1-5}
\hline
aa-9 & 143.4 \pm 4.7   &   38.1 \pm 5.3  &  5.1  \pm 0.3  &   186.6  \pm   10.3\\ \cline{1-5}
\hline
aa-10 & 145.9 \pm 2.2  &   38.2 \pm 1.0  &  4.4 \pm 0.3  & 188.5 \pm  3.5\\ \cline{1-5}
\hline
\end{tabular}
\end{table*}

The first three columns of the table describe the Coulomb, Lennard-Jones, Restraints contributions to free energy, respectively, while the last one corresponds to the value of the total ligand-protein complex free energy. All the values are expressed in $kJ\cdot mol^{-1}$. In Fig.~\ref{fig:_at_vs_dualres_fn_aa}, the atomistic reference is represented with a dash black line with its error bar. In particular, panels (a), (b) and (c) show the three components that contribute to the total complex free energy, reported in panel (d). Looking at these values as a function of the number of all-atom active site residues, we notice that there are important deviations of the free energy from the reference, especially in the case of 3 and 4 atomistic residues. On the contrary, the total value of the binding free energy agrees with the reference within the error bar in all cases.

Furthermore, we observe that the trend of free energy values, in comparison to the reference, is essentially the same: starting from 3 amino acids it approaches the reference until reaching 6, both in its components and in total. In contrast, going from 6 to 8 atomistic residues the value deviates from the reference, even though the total remains close to it. Finally, from 8 to 10, $\Delta G$ converges again. Hence, increasing the number of atomistic residues does not introduce necessarily an improvement of the computed free energy, at least as long as the various free energy components are considered separately.

\begin{figure}[t!]
\includegraphics[width=\columnwidth]{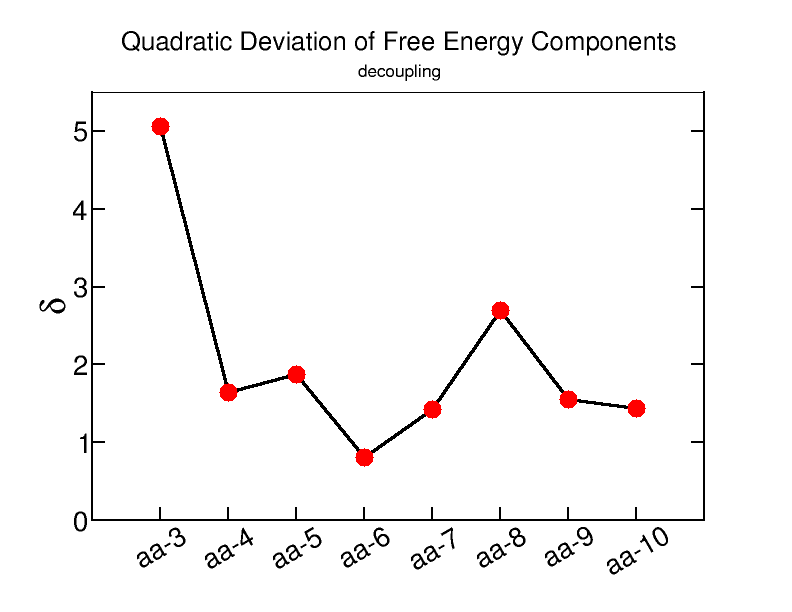}
\caption{Square root of quadratic deviation $\delta^2$ vs the number of atomistic residues chosen. The plot shows that in the case of 6 atomistic residues, the value of quadratic deviation is the lowest one and hence it means that such a number leads the best result of free energy. Moreover the black line shows the trend of FE values as discussed in the section \ref{res-disc} }.
\label{fig: sigma}
\end{figure}

In order to gain further, quantitative insight into these results, we computed the the quadratic deviation from the reference, $\delta^2$, defined as:
\begin{equation}\label{eq:stdev}
\begin{split}
\delta_{i}^2 &= \delta_{i-Coul}^2 + \delta_{i-LJ}^2 + \delta_{i-Restr}^2 =\\
& = (\Delta G_{Coul\_i}-\Delta G_{Coul-at})^2\\
& + (\Delta G_{LJ\_i}-\Delta G_{LJ-at})^2\\
& + (\Delta G_{Restr\_i}-\Delta G_{Restr-at})^2
\end{split}
\end{equation}
\medskip
where the index $i = 3 ... 10$ runs over atomistic residues. Fig.~\ref{fig: sigma} reports $\delta^2$ as a function of the number of active site amino acids modelled with atomistic detail.

The plot shows that the binding free energy computed in the dual-res model approaches the reference as the number of atomistic active site residues increases, and most importantly this approach takes place for each component up 6 residues. Beyond this value, though, the trend stops and the deviation becomes larger, peaking at 8 residues and decreasing when further atomistic amino acids are added. These results highlight a non-monotonic dependence of the free energy on the mapping, that is, the number of retained atomistic residues. If, on the one hand, the overall value of the binding free energy (Fig. \ref{fig:_at_vs_dualres_fn_aa} panel d) levels to the reference with as few all-atom residues as 4, the separate components oscillate and reach the plateau only for larger numbers. The existence of a minimum in the  standard deviation of all three contributions pinpoints a particular number of atomistic active site residues for which the accuracy of the computed free energy is the highest and the economy of the high-resolution subpart the largest. Including more than 6 atomistic residues counterintuitively worsens the result --when the various contributions are looked at-- and the previous accuracy is only recovered when more residues are included. This behaviour suggests that the total free energy undergoes an error cancellation which hides the deviations of the separate terms.

\begin{figure*}[t!]
\subfloat[]{\includegraphics[width=\columnwidth]{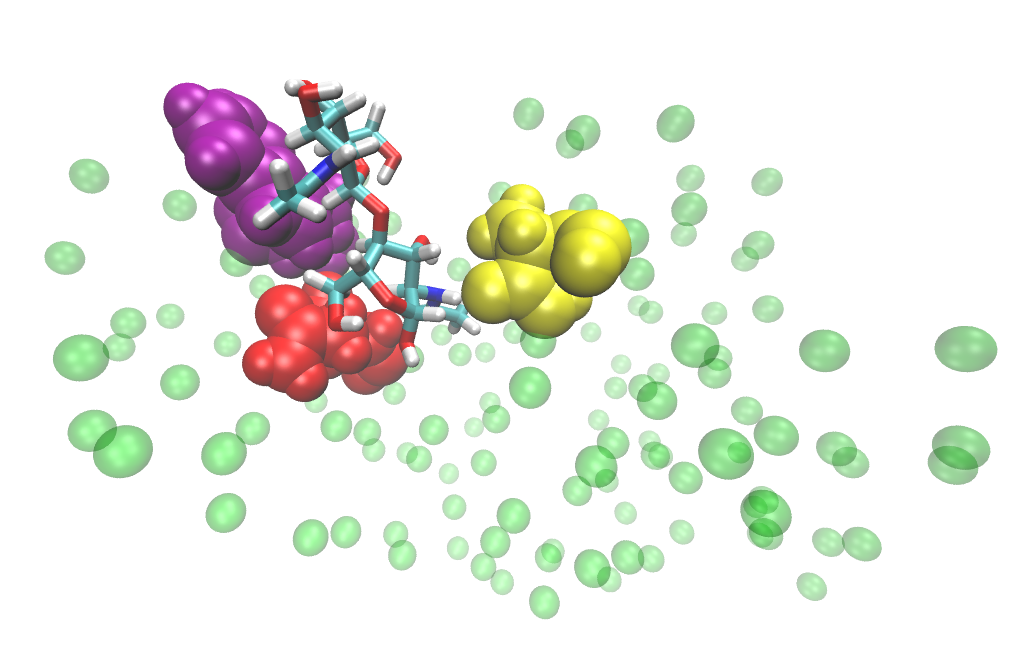}}
\subfloat[]{\includegraphics[width=\columnwidth]{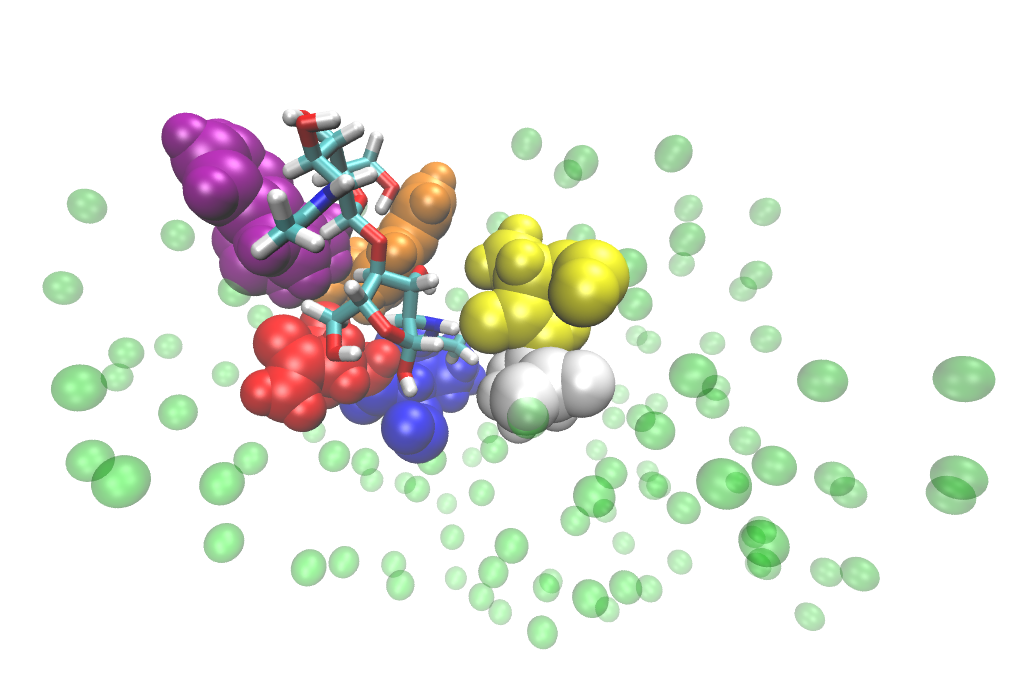}}\\
\subfloat[]{\includegraphics[width=\columnwidth]{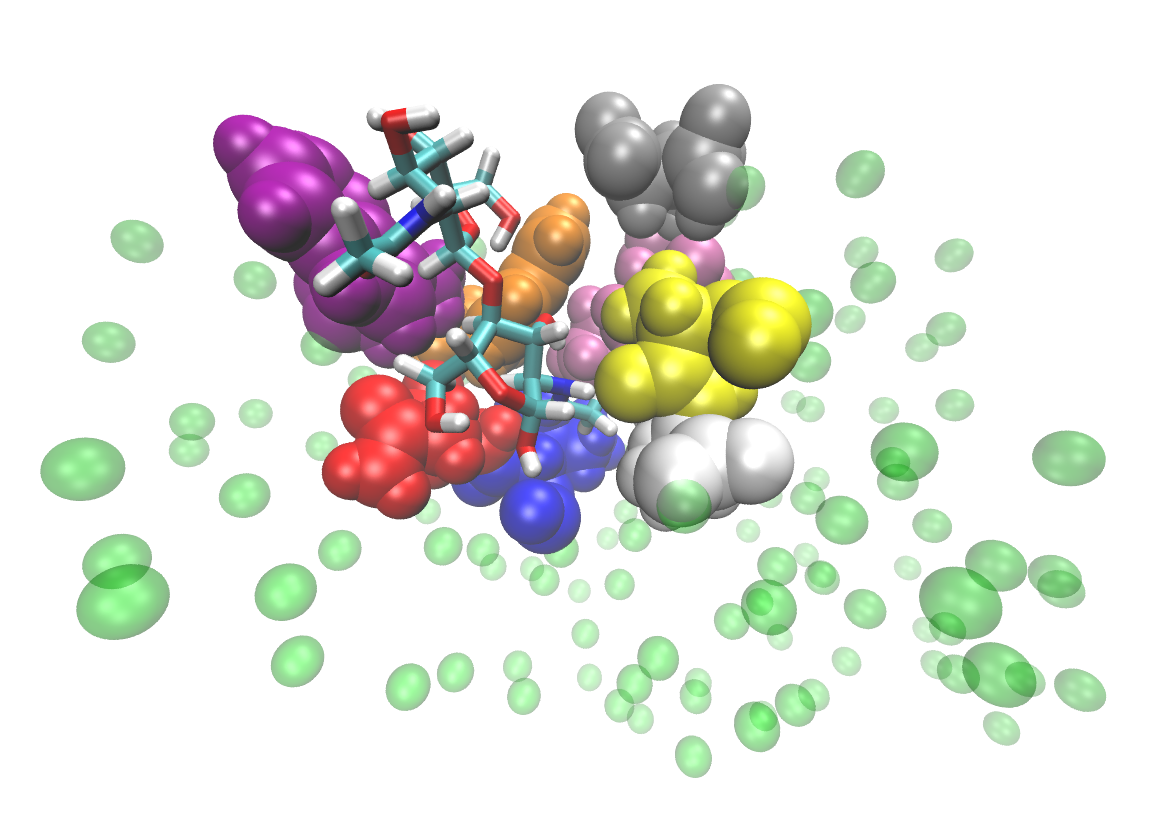}}
\subfloat[]{\includegraphics[width=\columnwidth]{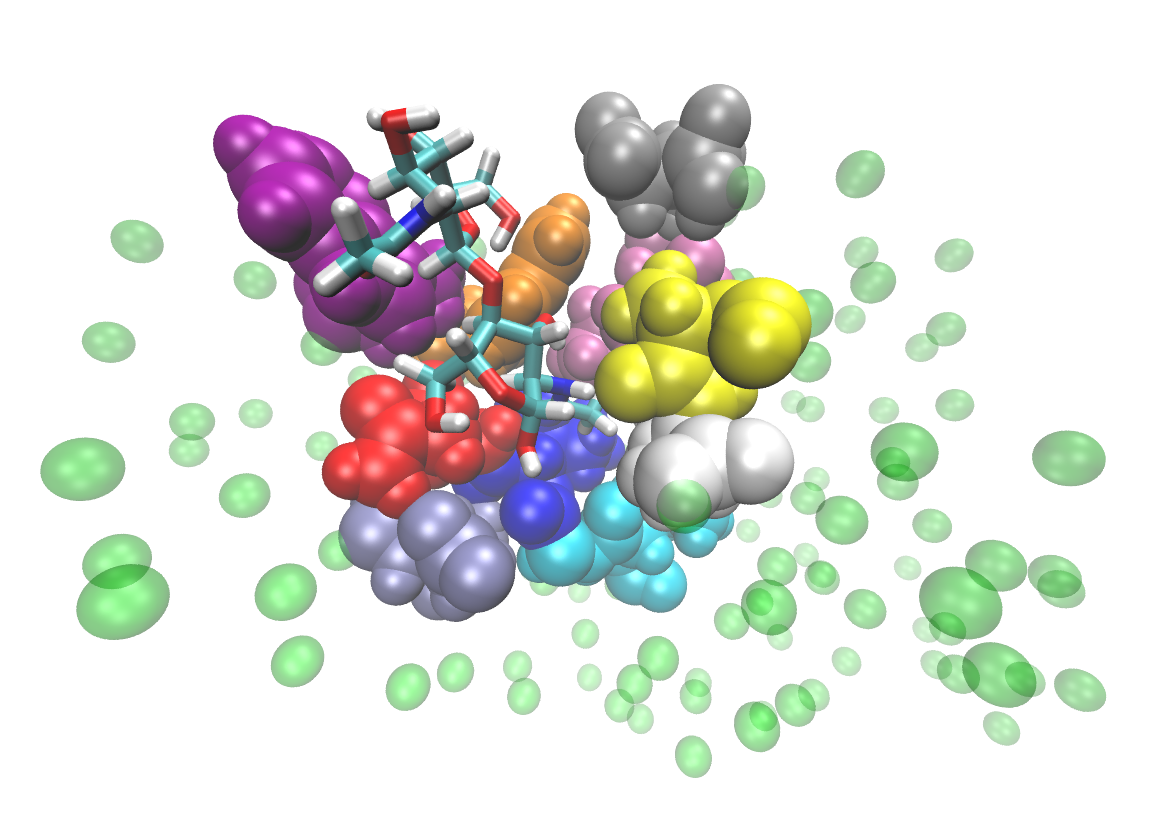}}\\
\caption{VMD representation of lysozyme and ligand in different resolution: (a) three, (b) six, (c) eight, (d) ten atomistic residues. The complete set can be found in \textit{Supporting information} The ligand is always atomistic and it is represented in Licorice. In green are represented the ENM beads. With the other colors are represented, instead, the various atomistic residues which surround the ligand}
\label{fig: ENM-at}
\end{figure*}

\begin{figure} [htp]
\includegraphics[width=\columnwidth]{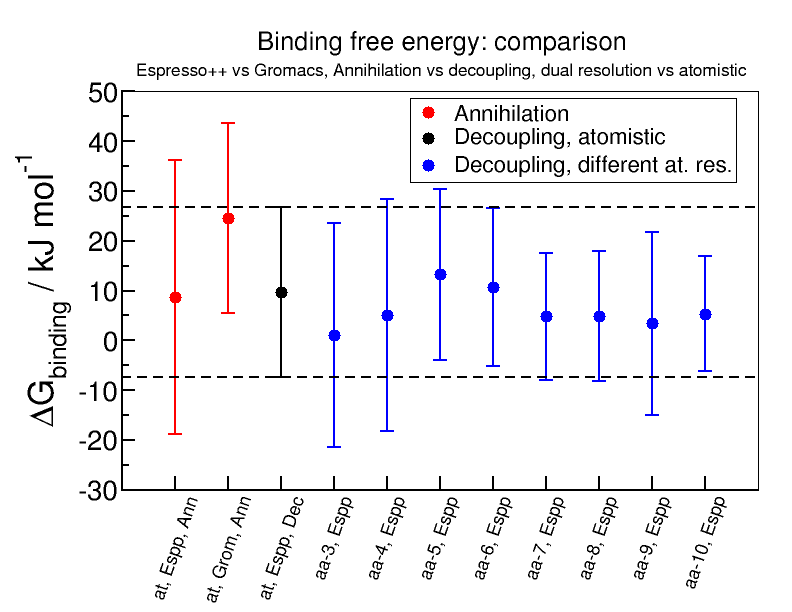}
\caption{Binding free energies as a function of protein's residues included in atomistic detail in the multi-resolution set-up or fully atomistic set-up. The heavy dashed black horizontal lines and black point are the reference values from fully atomistic simulations obtained in ESPResSo++ with decoupling, and the lighter dotted black horizontal lines are the error bars for those values. In red are represented binding free energies values in ESPResSo++ and GROMACS in case of annihilation. In blue is represented the binding FE value in dual resolution simulation changing the number of atomistic residues.}
\label{fig: comparison}
\end{figure}

A possible explanation for this nontrivial behaviour is that when 6 active site residues are modelled with all-atom accuracy (Fig.~\ref{fig: ENM-at}b) the ligand is stable in the catalytic site, namely it is surrounded by a complete shell of atomistic residues. The addition or deletion of other residues (Figs.~\ref{fig: ENM-at}c and \ref{fig: ENM-at}a respectively) leads to a worsening of $\Delta G$: in the first case, the two added residues (in pink and grey) are located behind the first shell of amino acids (far away from the ligand) and start to form a second, incomplete shell; in the second case, only three atomistic amino acids take part in the direct interaction with the ligand: therefore, the first layer is still incomplete and important interactions are missing; in order to improve the free energy value one has to add further amino acids in order to complete the second shell. We emphasise that the impact on the deviation from the reference is inversely proportional to the distance of the added/removed amino acid. Thus, the farther the atomistic amino acid is from the ligand, the more negligible its effect is. In the \textit{Supporting Information} we provide detail about the other numbers of all-atom residues not reported here. Finally, the values of binding free energy (also for the case of annihilation whose calculations are reported in the \textit{Supporting Information}) are summarised in Tab.~\ref{tab:finalres} and illustrated in Fig.~\ref{fig: comparison}.

\section{Conclusions}\label{concl}

In this work we have shown how the dual resolution model employed, constituted by an all-atom subregion coupled to an elastic network model remainder, can be used to calculate the binding free energy of an enzyme-substrate complex with atomistic accuracy. Furthermore, and most importantly, we have highlighted the impact that different choices of the model resolution can have. Specifically, we have computed the total value of the binding free energy as well as that of its various energetic components, and quantitatively inspected how these change when different selections are performed for the subgroup of amino acids, ranging from 3 to 10 in total, to be modelled at the fully atomistic level.

At first sight, one can appreciate that the binding free energy value rapidly converges to the atomistic reference when as few as 4 amino acids constituting the active site are described all-atom. This comforting result, however, unveils a greater complexity when the different terms constituting the free energy are looked at separately. These show an oscillating behaviour as the number of all-atom residues in the active site is increased, with a decreasing difference from the reference followed by a sudden jump to larger values, which dampens upon further addition of atomistic amino acids. The rationale in this behaviour is identified in the structure of the active site, which is constituted by a first shell of the six residues exposed to the solvent and closest to the ligand; when further amino acids beyond these are modelled with atomistic resolution, they interact with the substrate affecting the binding free energy components and shifting them away from the reference, with a steadily lowering impact as the model's resolution is increased - as one can expect. Surprisingly, very little if no signal of this behaviour is observed in the value of the binding free energy as a whole, rather it becomes visible only upon inspection of its separate contributions.

The results of this work thus highlight the importance of mapping in the construction of multi-scale and multi-resolution models, as a higher degree of detail does not necessarily correlate with a higher accuracy of the quantities of interest. The implications of these observations should serve as a warning in the realisation of coarse-grained models concurrently employing various levels of detail for different regions of the same system, whose range of application spans from fundamental understating of a molecule's properties to real-life pharmaceutical applications.

\acknowledgments{The authors are grateful to Robinson Cortes-Huerto and Thomas Tarenzi for a critical reading of the manuscript. This project has received funding from the European Research Council (ERC) under the European Union’s Horizon 2020 research and innovation programme (grant agreement no. 758588-VARIAMOLS and grant agreement no. 340906-MOLPROCOMP).}

\begin{table}[htpb]
\centering
\setlength{\tabcolsep}{0mm}
\caption{Representation of Free Energies values computed in ESPResSo++ and GROMACS (respectively \textit{espp} and \textit{grom} using a short notation on the table) in case of annihilation and decoupling. The table is divided in three column: from left to right are represented the ligand, protein-ligand complex and binding FE. The latter is the algebraic sum of $\Delta G_{compl}$, $\Delta G_{r\_off}$ and $\Delta G_{lig}$.  The results are in $kJ \cdot mol^{-1}$.}
\label{tab:finalres}
\begin{tabular}{@{}|l| 
                @{\hspace*{0mm}}  S[table-format=-5.1(4)]|
                @{\hspace*{0mm}}  S[table-format=5.1(4)]|
                @{\hspace*{0mm}}  S[table-format=3.1(4)]| 
                @{}}
\hline
& \textbf{Ligand}  & \textbf{Complex}  & \textbf{Binding}  \\
\hline
 \multicolumn{4}{|c|} {\textbf{annihilation}}  \\ \cline{1-4}  
\hline
  \textit{atom, espp  }   & -1275.3 \pm 11.2  &  1315.2 \pm 16.3   &  8.6 \pm 27.5\\ \cline{1-4}
  \textit{atom, grom  }   & -1259.0 \pm 5.9   &  1314.8 \pm 13.2  &  24.5 \pm 19.1\\ \cline{1-4}
  \hline
  \multicolumn{4}{|c|} {\textbf{decoupling}}  \\ \cline{1-4}
\hline
 \textit{atom, espp  }   & -151.9 \pm 8.0 & 193.0 \pm 9.1  &    9.8 \pm 17.1 \\ \cline{1-4}
\textit{aa-3, espp  } & -151.9 \pm 8.0 & 184.2 \pm 14.4  & 1.0 \pm 22.4  \\ \cline{1-4}
\textit{aa-4, espp  } & -151.9 \pm 8.0  & 188.3 \pm 15.3  & 5.1 \pm 23.3  \\ \cline{1-4}
\textit{aa-5, espp  } & -151.9 \pm 8.0 & 196.4 \pm 8.5  & 13.2 \pm 16.5  \\ \cline{1-4}
\textit{aa-6, espp  } & -151.9 \pm 8.0 & 193.8 \pm 7.8  & 10.6 \pm 15.8  \\ \cline{1-4}
\textit{aa-7, espp  } & -151.9 \pm 8.0  & 187.9 \pm 4.8  & 4.7 \pm 12.8  \\ \cline{1-4}
\textit{aa-8, espp  } & -151.9 \pm 8.0  & 188.0 \pm 5.1 & 4.8 \pm 13.1  \\ \cline{1-4}
\textit{aa-9, espp  } & -151.9 \pm 8.0  & 186.6 \pm 10.3 & 3.4 \pm 18.3   \\ \cline{1-4}
\textit{aa-10, espp  } & -151.9 \pm 8.0 & 188.5 \pm 3.5  & 5.3 \pm 11.5    \\ 
\hline
\end{tabular}
\end{table}

\bibliography{main}

\end{document}